\documentclass[12pt]{article}
\usepackage{latexsym}
\usepackage{amsmath}
\usepackage{wasysym}
\usepackage{amsfonts}
\usepackage{dsfont}
\usepackage{pifont}
\usepackage{bm}
\usepackage{epsf}
\usepackage{epsfig,graphicx}
\usepackage{amsmath,amssymb,bm,epsf,epsfig,graphicx}

\usepackage[vcentermath]{youngtab}

\usepackage[colorlinks=true,linkcolor=blue,citecolor=blue,urlcolor=black,filecolor=black,linktocpage=true]{hyperref}

 \csname
@addtoreset\endcsname{equation}{section}

\def\pe{\prime}
\def\3s{{s \choose 3}}
\def\4s{{s \choose 4}}
\def\5s{{s \choose 5}}
\def\6s{{s \choose 6}}

\def\12{\frac{1}{2}}
\def\fr{\frac}
\def\ft{\footnote}

\def\nn{\nonumber}

\def\pr{\partial}

\def\prd{\partial \cdot}

\def\be{\begin{equation}}
\def\ee{\end{equation}}
\def\bea{\begin{eqnarray}}
\def\eea{\end{eqnarray}}
\def\ba{\begin{array}}
\def\ea{\end{array}}
\def\bec{\begin{center}}
\def\ec{\end{center}}

\def\a{\alpha} 
\def\b{\beta}  
\def\g{\gamma} 
\def\G{\Gamma}
\def\d{\delta}

\def\h{\eta}

\def\L{\Lambda}
\def\m{\mu}

\def\r{\rho}

\def\vf{\varphi}

\def\cD{{\cal D}}

\def\cF{{\cal F}}

\def\cR{{\cal R}}

\begin{document}

\begin{center}


{\Large\sc Generalised connections and higher-spin equations}\\


\vspace{30pt} {\sc Dario Francia${}^{\; a,\,b}$} \\ \vspace{10pt} 
{${}^a$\sl\small Centro Studi e Ricerche E. Fermi\\
Piazza del Viminale 1, I-00184 Roma, Italy \\
${}^b$Scuola Normale Superiore and INFN \\ Piazza dei Cavalieri 7, I-56126 Pisa, Italy} \\
e-mail:
{\small \it dario.francia@sns.it }
\vspace{10pt}


\vspace{30pt} {\sc\large Abstract}\end{center}
We consider high-derivative equations obtained setting to zero the divergence of the higher-spin curvatures in metric-like form,  showing their equivalence to the second-order equations emerging from the tensionless limit of open string field theory, which propagate reducible spectra of particles with different spins. This result can be viewed as complementary to the possibility of setting to zero a single trace of the higher-spin field strengths, which yields an equation known to imply Fronsdal's equation in the compensator form. Higher traces and divergences of the curvatures produce a whole pattern of high-derivative equations whose systematics is also presented.

\vfill
\setcounter{page}{1}

\pagebreak




\section{Introduction}\label{sec:intro}


 The free propagation of massless particles of arbitrary spin can be described by means of one of the two equations 
\begin{align}
& \cF \, \equiv \, \Box \, \vf \, - \, \pr \, \prd \vf \, + \, \pr^{\, 2} \, \vf^{\, \pe} \, = \,  3 \, \pr^{\, 3} \, \a \, ,  \label{freeF} \\
& M \, \equiv \, \Box \, \vf \, - \, \pr \, \prd \vf \, = \, - \, 2 \, \pr^{\, 2} \, D \, , \label{freeM}
\end{align}
where  $\vf$ is a symmetric rank$-s$ tensor subject to the abelian gauge transformation $\d \, \vf \, = \, \pr \, \L$.  The first one is the Fronsdal equation \cite{fronsdal}, in its minimal unconstrained form with traceful parameter and compensator field $\a$ transforming as $\d \, \a \, = \, \L^{\, \pe}$ \cite{fs1, fs2}. It describes the propagation of a single massless particle of spin $s$. The second equation emerges from the tensionless limit of the free open string: it contains an additional field $D$ such that $\d \, D \, = \, \prd \L$, and propagates a whole set of massless particles with spin $s, \, s-2, \, s-4, \, \ldots$ and so on \cite{triplets1, triplets3, fs2, st, FoTsu}. Partial gauge fixings in (\ref{freeF}) and (\ref{freeM}) allow to eliminate the additional fields and reach the conventional Fronsdal form $\cF = 0$ for the first, with $\L^{\, \pe} = 0$, or the Maxwell-like equation $M = 0$ for the second, which is gauge invariant under transformations involving parameters whose divergence is constrained to vanish: $\prd \L = 0$ \cite{M}\ft{For general reviews on higher-spin theory see \cite{hsp1, hsp2, hsp3, hsp4}, while some alternative formulations of the free theory can be found in \cite{other1, other2, other3, other4, other6, other7}.}.

The Fronsdal tensor $\cF$ is identical in form to the linearised Ricci tensor of gravity, and one might wonder  how far this similarity can be pushed when it comes to investigating the geometric underpinnings of (\ref{freeF}). Indeed, one can obtain  $\cF$ computing the trace of a suitably defined second-order Christoffel-like connection, 
\be \label{connex}
\cF_{\, \m_1 \, \cdots \, \m_s} \, = \, \h^{\a \b}\, \G^{\, (2)}_{\, \a \b;\, \m_1 \, \cdots \, \m_s} \, ,
\ee
just as the linearised Ricci tensor is the trace of the corresponding Riemann curvature; however, for higher spins the role of linearised curvatures is better embodied by tensors of higher rank and higher derivative order and in this sense  (\ref{connex}) brings about both an additional analogy and an important difference between spin $2$ and higher spins, showing that Fronsdal's theory retains a different status if compared to its spin$-2$ precursor. Lagrangians for Fronsdal's particles exploiting higher-spin curvatures where proposed in \cite{fs1, fms}, and were generalized to the massive case in  \cite{dariomass, dariopropm}. With hindsight, a possible systematic procedure to generate them is to first remove Fronsdal's trace constraints --whose role appears unnatural from this vantage point-- by introducing additional fields like the compensator $\a$ in (\ref{freeF}) \cite{fs3}, and then integrate over the unphysical fields (instead of gauge-fixing them to zero), thus generating non-local, unconstrained actions amenable to be expressed in terms of higher-spin curvatures \cite{dariocrete}.  

  An alternative option leading to geometric, local, non-Lagrangian equations was explored in \cite{bb1, bb2} building on previous works \cite{damourdeser, henneauxdb} and exploiting in particular the generalized Poincar\'e lemma  \cite{henneauxdb, bb1}. (See also \cite{dmhull2}.) The idea is to define local equations mimicking the spin$-2$ case setting to zero a single trace of the higher-spin field strengths, to then show that the corresponding, high-derivative equations are indeed equivalent to Fronsdal's equation in its  unconstrained compensator form (\ref{freeF}):
\be \label{traceRintro}
\cR^{\, \pe}_{\, \r_{\, s-2}, \, \m_s} \, = \, 0 \hskip1cm \longrightarrow \hskip 1cm \cF \,  = \,  3 \, \pr^{\, 3} \, \a \, .
\ee
The purpose of this note is to investigate the counterpart of (\ref{traceRintro}) for the equation (\ref{freeM}), whose origin and meaning we shall briefly review in the following paragraphs.
 
 An alternative to Fronsdal's equation for the description of massless particles of arbitrary spin is provided by the equation 
\be \label{maxwell}
M \, = \, \Box \, \vf \, - \, \pr \, \prd \vf \, = \, 0 \, , \\
\ee
a formal extension to all spins of  Maxwell's theory where  gauge invariance obtains requiring that the gauge parameter be divergence-free: $\prd~\L = 0$ \cite{M}. Equation (\ref{maxwell}) describes the propagation of reducible multiplets of massless particles, with spins ranging in the set $s, s-2, s-4, \ldots$ down to $s=1$ or $s=0$. However, adding the condition that the gauge field $\vf$ be traceless --in this case the first equation in (\ref{maxwell}) must also be modified and changed to its traceless projection--  it is also possible to truncate this multiplet to the single representative of highest spin \cite{sv}.  Moreover, as an additional analogy with its spin$-1$ precursor, the Maxwell tensor $M$ for any spin can be obtained computing a single divergence of a suitably defined first-order connection $\G^{\, (1)}_{\, \r, \, \m_1 \, \cdots \, \m_s}$:
\be \label{Tdiff}
M_{\, \m_1 \, \cdots \m_s} \, = \, \pr^{\, \r} \, \G^{\, (1)}_{\, \r, \, \m_1 \cdots  \m_s} \, .    
\ee

The tensionless-string Lagrangians leading to (\ref{freeM}) provide somehow the minimal unconstrained extensions of (\ref{maxwell}), and as such can be exploited to establish a more direct link to higher-spin curvatures. Indeed, instead of eliminating the compensator field $D$ by a partial gauge fixing one can integrate it away, obtaining non-local Lagrangians as simple as squares of higher-spin field strengths and leading to equations of motion only involving divergences of the curvatures  \cite{dariotripl}. In view of the analogies  between the constructions of irreducible and reducible theories, in \cite{dariotripl, dariokyoto}  the existence of a local, high-derivative counterpart of (\ref{traceR}) was also surmised, and it was suggested that it should take the form of a transversality condition on the curvature itself:
\be \label{divRintro}
\prd \cR_{\, \r_{\, s-1}, \, \m_s} \, = \, 0 \hskip1cm \longrightarrow \hskip 1cm M \,  = \,  -\,2 \, \pr^{\, 2} \, D \, .
\ee
The main goal of this note is thus to provide a proof of this statement. As an additional output of our investigation we shall also observe that the two correspondences (\ref{traceRintro}) and (\ref{divRintro}) actually define the simplest instances of a whole hierarchy of possible equations involving  traces and divergences of higher-spin curvatures. The tensors obtained taking successive traces of the curvatures --together with possibly a single divergence-- and those arising after computing a number of divergences, coincide, up to multiplication by the D'Alembertian operator, with the class exploited in \cite{fs1, dariotripl} in the construction of unconstrained Lagrangians\ft{An analogous construction is also considered in \cite{EK}, providing the building-blocks for actions invariant under generalized Weyl symmetry.}. However, a whole class of kinetic operators stemming out from our analysis is new and we leave  a more detailed evaluation of their possible dynamical role to future work.

We shall first review the construction of higher-spin connections and curvatures following \cite{dwf}, to then discuss in section 3 the reduction of the equations obtained setting to zero  their divergence or their trace. Finally, in section 4 we shall consider the generalized pattern of high-derivative equations whose corresponding kinetic tensors transform with higher powers of gradients of traces and divergences of the parameter. For our notation and conventions the reader can consult \cite{fs1, dariomass}.

\section{Higher-spin curvatures} \label{sec:curv}

Linearised curvatures for gauge potentials of arbitrary rank can be defined as the simplest tensors that are identically gauge invariant under 
\be
\d \, \vf \, = \, \pr \, \L \, ,
\ee
and do not vanish on-shell, i.e. when $\vf$ satisfies its free equations of motion.

Metric-like tensors with these properties were defined in \cite{weinberg, dwf, damourdeser}. We refer in particular to  the  construction given by de Wit and Freedman in \cite{dwf}, in which the higher-spin field strengths emerge as  distinguished elements of a whole hierarchy of generalised  ``Christoffel connections'': the first element in the hierarchy (of zero order in derivatives) can be identified with the gauge potential itself, while higher-order connections are built iteratively taking successive derivatives of the gauge field. 
Concretely:
\be \label{connexions}
\begin{split}
&\G^{(0)}_{\, \m_{s}} \, = \, \vf_{\, \m_s} \, , \\
&\G^{(1)}_{\, \r,\,  \m_{s}} \, = \, \pr_{\, \r}\, \vf_{\, \m_s} \, - \, \pr_{\, \m} \, \vf_{\, \r \, \m_{s-1}}, \\
&\G^{(2)}_{\, \r_{2},\,  \m_{s}} \, = \, \pr^{\,2}_{\, \r}\, \vf_{\, \m_s} \, - \, \12 \, \pr_{\, \r}\, \pr_{\, \m} \, \vf_{\, \r \, \m_{s-1}} \, 
+ \, \pr^{\, 2}_{\, \m} \, \vf_{\, \r_{2} \m_{s-2}}\, , \\
& \, \ldots \, , \\
& \G^{(n)}_{\, \r_{n},\, \m_{s}} =   \ \sum_{k=0}^{m}\fr{(-1)^{k}}{\left({{n} \atop {k}} \right)}\ 
\pr^{\, n-k}_{\, \r}\, \pr^{\, k}_{\, \m} \, \vf_{\, \r_{k} \, \m_{s-k}}\, , 
\end{split}
\ee
where $m \, \equiv min\, (n, \, s)$. The corresponding gauge transformations are
\be \label{bosetrans}
\begin{split}
&\d\, \G^{(n)}_{\, \r_{n},\, \m_{s}} \, = \,(-1)^n \, (n+1)\,  \pr^{\, n+1}_{\, \m} \, \L_{\, \r_{n},\, \m_{s-n-1}}\, , \hskip .5cm n \, < \, s\\
&\d\, \G^{(n)}_{\, \r_{n},\, \m_{s}} \, = \, 0 \, , \hskip .5cm n \, \geq \, s\, .
\end{split}
\ee
Let us observe that the tensors $\G^{(n)}_{\, \r_{n},\, \m_{s}}$ are in general reducible. In particular,  for $0\, \leq \,n\, \leq s$ the only  $GL(D)-$irreducible tensor besides the field $\vf$ itself is the $s-$th element of the hierarchy, that one identifies as the proper candidate to play the role of a higher-spin curvature
\be \label{curvature}
\cR_{\, \r_s, \,  \m_s} \, = \, \sum_{k=0}^{s}\fr{(-1)^{k}}{\left({{s} \atop {k}}\right)}\ 
\pr^{s-k}_{\, \r}\pr^{k}_{\, \m} \vf_{\, \r_k,\,\m_{s-k}} \, . 
\ee
For the full set of connections generalised Bianchi identities hold,
\be
 \pr_{[\a} \, \G^{(n)}{}_{\, \b]\, \r_{m-1}, \, \g \, \m_{s-1}}\, + \, \pr_{[\g} \, \G^{(n)}{}_{\, \a]\, \r_{m-1}, \, \b \, \m_{s-1}}\,
+ \, \pr_{[\b} \, \G^{(n)}{}_{\, \g]\, \r_{m-1}, \, \a \, \m_{s-1}}\, \equiv \, 0 \, ,
\ee
while the tensors in (\ref{curvature}) also satisfy 
\be \label{irr}
\cR_{\, \r_{s-1} \, \m, \,  \m_s} \, =\, 0 \, , \hskip 1cm \cR_{\, \r_s, \,  \m_s} \, = \, (-1)^{\, s} \, \cR_{\, \m_s, \,  \r_s} \, .
\ee
From (\ref{bosetrans}) it is clear that the traces of the connections $\G^{(n)}_{\, \r_{n},\, \m_{s}}$, for $n< s$, computed contracting two $\r-$indices, transform proportionally to traces of the gauge parameter, while for divergences taken within the same group divergences of the parameters would appear. Thus, gauge invariance extends to the first traces of all  tensors in (\ref{connexions}) with $n \geq 2$ at the price of assuming that the gauge parameter be traceless, consistently in particular with the identification of the Fronsdal tensor with the trace of $\G^{(2)}_{\, \r_{2},\, \m_{s}}$, as indicated in (\ref{connex}), while similar considerations lead to identify the Maxwell-like tensor $M$ with the divergence of  $\G^{(1)}_{\, \r,\, \m_{s}}$. Let us mention that the construction of the system \eqref{connexions} applies equally well to spinor-tensors and might serve as a starting point for a geometric description of fermionic dynamics \cite{fs1, fms, dariomass, dariotripl}.

\section{High-derivative equations from curvatures }

 The main goal of this note is to investigate the meaning of the equation
\be \label{div}
\pr^{\, \a} \, \cR_{\, \a \, \r_{s-1}, \,  \m_s} \, = \, 0 \, ,
\ee
obtained setting to zero the divergence of the higher-spin curvature  (\ref{curvature}). We shall show that, despite being a high-derivative equation,  (\ref{div}) is actually equivalent to the second-order equation  (\ref{freeM}),
\be \label{div1}
\prd \cR_{\,  \r_{s-1}, \,  \m_s} \,= \, 0 \hskip .5cm \iff \hskip .5cm  M_{\, \m_s} \, + \, 2 \, \pr^{\, 2}_{\, \m} \, D_{\, \m_{\, s-2}} \, = \, 0 \, ,
\ee
and thus describes the propagation of a multiplet of particles with different spins. The argument presented here easily extends to provide an alternative proof of (\ref{traceRintro}), that  we shall also present  in parallel with (\ref{div1}). For a related discussion in the context of quantum forms on K\"ahler spaces see \cite{iazbas}.

 The first step is to compute a divergence or a trace of (\ref{curvature}) and recognize the emergence of the tensor $M$ in the former case and of the Fronsdal tensor $\cF$ in the latter:
\begin{align}
\prd \cR_{\,  \r_{s-1}, \,  \m_s} & = \, \sum_{k=0}^{s-1}\fr{(-1)^{k}}{\left({{s} \atop {k}}\right)}\ 
\pr^{s-k-1}_{\, \r}\pr^{k}_{\, \m} \, M_{\, \r_k,\,\m_{s-k}}\, , \label{divR} \\
\cR^{\, \pe}_{\, \r_{s-2}, \,  \m_s} & =  \, \sum_{k=0}^{s-2}\fr{(-1)^{k}}{\left({{s} \atop {k}}\right)}\ 
\pr^{s-k-2}_{\, \r}\pr^{k}_{\, \m} \,  \cF_{\, \r_k,\,\m_{s-k}} \, . \label{traceR} 
\end{align}
Under $\d \, \vf \, = \, \pr \, \L$ the tensors $M$ and $\cF$ transform as follows
\be
\d \, M \, = \, - \, 2 \, \pr^{\, 2} \, \prd \L \, , \hskip 1cm \d \, \cF \, = \, 3 \, \pr^{\, 3} \, \L^{\, \pe} \, ,
\ee
so that the gauge invariance of the curvatures makes it manifest that the  equations 
\begin{align} 
\prd \cR_{\,  \r_{s-1}, \,  \m_s} & = \, 0\, , \label{geomeqdiv} \\
\cR^{\, \pe}_{\, \r_{s-2}, \,  \m_s} & =  \, 0 \, , \label{geomeqtrace} 
\end{align}
are satisfied if the tensors $M$ and $\cF$ satisfy the corresponding unconstrained equations (\ref{freeF}) and (\ref{freeM}) for arbitrary tensors $D$ and $\a$, and the issue at stake is to show that there are no other solutions. The basic idea is to exploit the irreducibility of the curvatures in order to first identify those components of $M$ and $\cF$ that are left arbitrary by the corresponding equations and that eventually turn out to play the role of the compensators $D$ and $\a$. The second part of the proof consists in showing that the remaining components of $M$ and $\cF$ are linked to those arbitrary ones according to the corresponding equations. In the following we shall work in momentum space in light-cone coordinates, choosing a reference frame where $p_{\, +} \, \neq \, 0$. In order not to complicate the notation we shall denote Fourier coefficients with the same symbols as the corresponding space-time tensors, so that e.g.  $M \, (p) \, \equiv \, M$.

\subsection{$\prd \cR_{\,  \r_{\, s-1}, \, \m_s}  = \, 0$}

 The starting point for our discussion is to observe that the divergence of the curvature (\ref{divR}) defines an irreducible $GL(D)-$ tensor,
\be \label{s1}
\prd \cR_{\,  \r_{s-1}, \,  \m_s} \hskip .3cm \longrightarrow \hskip .3cm
\overbrace{\young(\hfil \hfil \ldots \hfil \hfil,\hfil \hfil \ldots \hfil)}^{s} 
\ee
as can be easily verified making use of (\ref{irr}). Because of this property, its components with all indices in the $\m-$group together with at least one index in the $\r-$group taking the same value vanish identically, e.g.
\be
\prd \cR_{\,  \r_{s-2}\, +, \, +_s} \, \equiv \, 0 \, .
\ee
As a consequence, because of (\ref{divR}),  the equations in  (\ref{geomeqdiv}) only involving components of the form $M_{\m_k \, +_{\, s-k}}$, with $k = 0 \, \cdots \, s-2$,  reduce to identities. These components in their turn can be encoded in an arbitrary rank$-(s - 2)$ tensor $\hat{D}_{\m_{s-2}}$  defining
\be \label{Ddef}
M_{\m_k \, +_{\, s-k}} \, = \, p^{\, 2}_{\, \m} \, \hat{D}_{\, \m_{\, k-2} \, +_{s - k}} \, + \, (s - k)\, p_{\, \m} p_{\, +} \, \hat{D}_{\, \m_{\, k-1} \, +_{\, s-k-1}} \, + \, {s - k \choose 2} \,  p^{\, 2}_{\, +} \, \hat{D}_{\, \m_{\, k} \, +_{s - k - 2}} \, .
\ee
What we would like to prove is that  (\ref{div}) establishes a link among the remaining components of $M$ and those in (\ref{Ddef}) taking the same form as  (\ref{freeM}). To this end, let us write the condition of transversality of the curvature for the set of components $(\r_{\, s-1}, \, \m_{\, s}) \, = \, (+_{\, s-1}, \, \m_{\, s})$:
\be
\begin{split}
p \cdot \cR_{\,  +_{s-1}, \,  \m_s} & = \, \sum_{k=0}^{s-1}\fr{(-1)^{k}}{\left({{s} \atop {k}}\right)}\, {\left({{s -1} \atop {k}}\right)} \, p^{s-k-1}_{\, +}\, p^{k}_{\, \m} \, M_{\, +_k,\,\m_{s-k}}\, \\
& = \, \sum_{k=0}^{s-1} \, (-1)^{k}\, \fr{s - k}{s} \, 
p^{s-k-1}_{\, +}\, p^{k}_{\, \m} \, M_{\, +_k,\,\m_{s-k}}\, = \, 0 \, ,
\end{split}
\ee
and then solve for the first term in the sum, obtaining
\be \label{M}
M_{\, \m_s} \, + \, \sum_{k=1}^{s-1} \, (-1)^{k}\, \fr{s - k}{s} \, 
\left(\fr{p_{\, \m}}{p_+}\right)^{k} \, M_{\, +_k,\,\m_{s-k}}\, = \, 0 \, ,
\ee
for arbitrary indices $\m_1 \, \ldots \, \m_s$. From the last formula one can first solve for $M_{\, + \m_{s-1}}$ getting
\be
M_{\, + \m_{s-1}} \, = \, \sum_{k=0}^{s-2} \, (-1)^{k}\, \left(\fr{p_{\, \m}}{p_+}\right)^{k+1} \, M_{\, +_{k+2},\,\m_{s-k-2}}\, ,
\ee
to then express $M_{\, \m_s}$ as a function of the components with at least two $``+''$ indices as follows:
\be
M_{\,\m_s} \, = \, \sum_{k=0}^{s-2} \, (-1)^{k}\, (k + 1) \, \left(\fr{p_{\, \m}}{p_+}\right)^{k+2} \, M_{\, +_{k+2},\,\m_{s-k-2}}\, .
\ee
Taking the definitions (\ref{Ddef}) into account  one can finally put the resulting equation in the desired form,
\be
M_{\, \m_{\, s}} \,  = \, p^{\, 2}_{\, \m} \, \hat{D}_{\, \m_{\, s-2}} \, .
\ee
%

\subsection{$\cR^{\, \pe}_{\,  \r_{\, s-2}, \, \m_s}  = \, 0$}

 In the same fashion as for the divergence, also the trace of the higher-spin curvature (\ref{curvature}) defines an irreducible $GL(D)-$ tensor,
\be \label{s2}
\cR^{\, \pe}_{\,  \r_{s-2}, \,  \m_s} \hskip .3cm \longrightarrow \hskip .3cm
\overbrace{\young(\hfil \hfil \ldots \hfil \hfil \hfil,\hfil \hfil \ldots \hfil)}^{s} 
\ee
so that, in particular, all its components with indices in the $\m-$group together with at least one index in the $\r-$group along the same direction, say the $+$ axis, vanish identically,
\be
\cR^{\, \pe}_{\,  \r_{s-3}\, +, \, +_s} \, \equiv \, 0 \, .
\ee
As a  consequence, the equations in (\ref{geomeqtrace}) for  components of $\cF$ with at least three $+$ indices, i.e. $\cF_{\m_k \, +_{\, s-k}}$, with $k = 0 \, \cdots \, s-3$ are identically satisfied.  Concretely it is useful to consider (\ref{geomeqtrace}) with all $\r$ indices along the $+$ direction,
\be
\cR^{\, \pe}_{\, +_{s-2}, \,  \m_s} \, =  \, \sum_{k=0}^{s-2} \,(-1)^k \, \fr{(s - k)\,(s-k-1)}{s\, (s-1)} 
p^{s-k-2}_{\, +}\, p^{k}_{\, \m} \,  \cF_{\, +_k,\,\m_{s-k}} \, = \, 0 \, ,
\ee
which, in its turn, implies an equation for the Fronsdal tensor of the form
\be \label{F}
\cF_{\, \m_s} \, = \, \sum_{k=1}^{s-2} \,(-1)^{k+1} \, \fr{(s - k)\,(s-k-1)}{s\, (s-1)} \, 
\left(\fr{p_{\, \m}}{p_{\, +}}\right)^{\, k} \,  \cF_{\, +_k,\,\m_{s-k}} \, ,
\ee
which subsumes all implications of (\ref{geomeqtrace}). In particular, while being an identity for all components of $\cF$ with at least three $+$ indices, (\ref{F}) relates the latter components to the remaining ones according to 
\be \label{F+}
\cF_{\, \m_s} \, = \, \sum_{k=1}^{s-2} \,(-1)^{k+1} \, {k + 1 \choose 2} \, 
\left(\fr{p_{\, \m}}{p_{\, +}}\right)^{\, k+2} \,  \cF_{\, +_{k+2},\,\m_{s-k-2}} \, ,
\ee
so that the Fronsdal tensor is equated to the cubic gradient of an arbitrary tensor. To establish a closer contact with (\ref{freeF}) one can redefine the components of  $\cF_{\, +_{\, k+2} \, \m_{s-k-2}}$ for  $k = 1 \, \cdots \, s-1$ introducing a rank$-(s-3)$ tensor $\a_{\, \m_{s-3}}$ as follows
\begin{eqnarray} \label{alpha}
\cF_{\, +_{\, k + 2} \, \m_{s-k -2}} & \equiv \, 3 \, \bigg(p_{\, \m}^{\, 3} \, \a_{\, +_{\, k + 2} \, \m_{s-k -5}}\, 
+ \, (k\, + \, 2)\, p_{\, \m}^{\, 2} \, p_+ \, \a_{\, +_{\, k + 1} \, \m_{s-k -4}} \,  \nn \\
&  + \, {k+2 \choose 2} \, p_{\, \m} \, p^{\, 2}_+ \, \a_{\, +_{\, k} \, \m_{s-k -3}} \, 
+ \, {k+2 \choose 3}\, p^{\, 3}_+ \, \a_{\, +_{\, k -1} \, \m_{s-k -2}} \bigg) \, ,   
\end{eqnarray}
from which, after substituting (\ref{alpha}) in (\ref{F+}), one finds 
\be
\cF_{\, \m_s} \, = \, 3 \, p^{\, 3}_{\, \m} \, \a_{\, \m_{s-3}} \, ,
\ee
as we wanted to prove.

\section{General pattern of high-derivative equations}

In this section we would like to show that (\ref{traceRintro}) and (\ref{divRintro}) are special members of a class of relations involving high-derivative combinations of $\cF$ and $M$ themselves. In order to get an intuition of the underlying pattern it is useful to recall that, according to their defining property, the curvatures (\ref{curvature}) vanish if and only if the corresponding potential is pure gauge  \cite{damourdeser} and then compare this property with the results discussed in the previous section:
\begin{align} \label{com1}
&& \cR_{\, \r_s, \, \m_s} \, & = \, \sum_{k=0}^{s}\fr{(-1)^{k}}{\left({{s} \atop {k}}\right)}\ 
\pr^{\, s-k}_{\, \r}\, \pr^{\, k}_{\, \m} \, \vf_{\, \r_k,\,\m_{s-k}} \, = \, 0 \, 
\hskip .2cm && \iff && \hskip .2cm \, \vf \, = \, \pr \, \L \, , \nonumber \\
&& \prd \cR_{\, \r_{s-1}, \, \m_s} \, & = \, \sum_{k=0}^{s-1}\fr{(-1)^{k}}{\left({{s} \atop {k}}\right)}\ 
\pr^{\, s-k-1}_{\, \r}\, \pr^{\, k}_{\, \m} \, M_{\, \r_k,\,\m_{s-k}} \, = \, 0 \, 
\hskip .2cm && \iff && \hskip .2cm \, M \, = \, - \, 2 \, \pr^{\, 2} \, D \, ,\\
&& \cR^{\, \pe}_{\, \r_{s-2}, \, \m_s} \, & = \, \sum_{k=0}^{s-2}\fr{(-1)^{k}}{\left({{s} \atop {k}}\right)}\ 
\pr^{\, s-k-2}_{\, \r}\, \pr^{\, k}_{\, \m} \, \cF_{\, \r_k,\,\m_{s-k}} \, = \, 0 \, 
\hskip .2cm && \iff && \hskip .2cm \, \cF \, = \,  3 \, \pr^{\, 3} \, \a \, . \nonumber
\end{align}
Let us also stress that, insofar as getting the last column in (\ref{com1}) is concerned, not only $\vf$ but also $M$ and $\cF$ can be regarded as generic symmetric tensors. In a sense, the sums in the second and third equations in (\ref{com1}) build the curvatures for gauge ``potentials'' transforming as double or triple gradients of arbitrary parameters, similarly to what the sum in the first equation does for the actual gauge potential $\vf$. Repeating the arguments of the previous section one can then recognize the existence of a general pattern of similar relations of the form
\be \label{com2}
\sum_{k=0}^{s-t}\fr{(-1)^{k}}{\left({{s} \atop {k}}\right)}\, 
\pr^{\, s-k-t}_{\, \r}\, \pr^{\, k}_{\, \m} \, \cD^{\, (t)}_{\, \r_k,\,\m_{s-k}} \, = \, 0 
\hskip .2cm  \iff  \hskip .2cm \, \cD^{\, (t)} \, = \, \pr^{\, t + 1} \, \Delta^{\, (t)}\, ,
\ee
for any symmetric, rank-$s$ tensor $\cD^{\, (t)}$, with arbitrary rank$-(s-t-1)$ tensors $ \Delta^{\, (t)}$ \cite{henneauxdb}. 

 It can also be interesting to investigate the possible explicit realizations of the tensors $ \cD^{\, (t)}$ as kinetic tensors for the ordinary potential $\vf$; they should emerge naturally computing higher traces and divergences of the curvatures (\ref{curvature}), while the structure of their gauge transformations, involving a given number of traces and divergences of the parameter, should match the second equation in (\ref{com2}). Roughly speaking, one expects that computing  $m$ divergences and $n$ traces of the curvature would give rise to a combination of gradients of a kinetic tensor transforming proportionally to $m$ divergences and $n$ traces of the gauge parameter. Setting to zero the corresponding tensor would thus give rise to an equation of the form (\ref{com2}) with $t = 2n + m$.
 
  Let us denote the corresponding solution for  $\cD^{\, (2n + m)}$  as $\cF^{\, (n, \, m)}$, where $n$ and $m$ refer to the numbers of corresponding traces and divergences of $\cR$, respectively. Schematically:
\be
\begin{split}
& \prd^{m} \cR^{\, [n]} \,  = \, \sum_{k=0}^{s-2n - m}\fr{(-1)^{k}}{\left({{s} \atop {k}}\right)} 
\,\pr_{\, \r}^{\, s - k - 2n - m}\, \pr_{\, \m}^{\, k}\, \cF^{\, (n, \, m)}_{\, \r_{\, k} \, \m_{\, s-k}} \, \\
& \d \, \cF^{\, (n, \, m)} \, \sim  \, \prd^m \L^{\, [n]} \, ,
\end{split}
\ee
where $\vf$, $M$ and $\cF$ are easily identified as the first elements of this double sequence: 
\begin{align} \label{initialdata}
& \cF^{\, (0, 0)} = \vf
\hskip .2cm && \longrightarrow && \hskip .2cm \, \d \, \vf \, = \, \pr \, \L \, , \nonumber \\
& \cF^{\, (0, 1)}  =  M  
\hskip .2cm && \longrightarrow && \hskip .2cm \, \d \, M \, = \, - \, 2 \, \pr^{\, 2} \, \prd \L \, ,\\
& \cF^{\, (1, 0)}  =  \cF 
\hskip .2cm && \longrightarrow && \hskip .2cm \, \d \, \cF \, = \,  3 \, \pr^{\, 3} \, \L^{\, \pe} \, . \nonumber
\end{align}
The complete systematics can then be uncovered via an iterative procedure. In the resulting system one can highlight the special classes of operators obtained computing only divergences  (``pure Maxwell-like''),
\be \label{pureM}
\begin{split}
&\cF^{\, (0, \, m)} \,  =  \, \Box  \, \cF^{\, (0, m-1)} \, - \, \fr{1}{m} \, \pr \, \prd \cF^{\, (0, m-1)},  \\
&\d \, \cF^{\, (0, m)} \, = \, (-1)^{m} \, (m \,+ \,1) \, \pr^{\, m + 1} \, \prd^m \L  , 
\end{split}
\ee
or only traces (``pure Ricci-like''),
\be \label{pureF}
\begin{split}
&\cF^{\, (n, \, 0)} \,  =  \, \Box  \, \cF^{\, (n-1, 0)} \, - \, \fr{1}{n} \, \pr \, \prd \cF^{\, (n-1, 0)} \, + \, 
\fr{1}{n\, (2n -1)} \, \pr^{\, 2} \,  \cF^{\, (n-1, 0) \, \pe}  \\
&\d \, \cF^{\, (n, 0)} \, =  \, (2 n \,+ \,1) \, \pr^{\, 2 n + 1} \, \L^{\, [n]}  \, ,
\end{split}
\ee
while in order to exhibit the general solution it is useful to proceed computing successive divergences of $\cR^{\, [n]}$,  which leads to 
\be \label{general}
\begin{split}
&\cF^{\, (n, \, m)}  \, = \, \Box  \cF^{\, (n, m-1)}  \, - \,  \fr{1}{2n  +  m} \,  \pr \,  \prd \cF^{\, (n, m-1)} \, , \\
&\d \, \cF^{\, (n, \, m)} \, = (\, -1)^{\, m} \,  (2 n + m + 1) \, \pr^{\, 2 n + m + 1} \,  \prd^m \L^{\, [n]} \, .
\end{split}
\ee
Let us also mention that the kinetic tensors in (\ref{pureF}) actually coincide (up to a multiplication by the D'Alembertian operator) with the sequence of geometric kinetic tensors obtained in \cite{fs1} and there denoted by $\cF^{\, (n)}$, while those in (\ref{pureM}) emerged in \cite{dariotripl} after integrating away the field $D$ from (\ref{freeM}). Gauge fixing to zero the generalized compensators $ \Delta^{\, (t)}$ leads to constrained, higher-derivative equations some of which, corresponding to the classes $\cF^{\, (n, \, 0)}$ and $\cF^{\, (n, \, 1)}$, are also investigated in \cite{EK} in the context of possible high-derivative actions with generalized constrained symmetries.


\section{Conclusions}

 The higher-spin curvatures of \cite{weinberg, dwf, damourdeser} provide a long-recognized elegant structure, more recently shown to be capable of encoding all relevant information concerning free higher-spin dynamics. However, their remarkable algebraic properties notwithstanding, the role of metric-like higher-spin curvatures in clarifying features of higher-spin interactions is at present little explored --aside from the trivial possibility of making use of them to define abelian-invariant vertices \cite{damourdeser2}-- while their non-linear deformations are known only up to the quadratic contribution computed in \cite{manvR}.  It is anyway noteworthy that one can formulate high-derivative equations propagating only the modes of the  two-derivative equations defined in (\ref{freeF}) and (\ref{freeM}). High derivatives unavoidably appear in higher-spin interactions even when the kinetic term is written in its ordinary, two-derivative form, and this poses in general an issue for what regards the proper formulation of the corresponding Cauchy problem. In this sense, it would be interesting to investigate whether a mechanism  similar to the one here highlighted for the free theory might be of help in clarifying this aspect of higher-spin dynamics. 
  

\section*{Acknowledgments}

I would  like to thank X. Bekaert, N. Boulanger and K. Mkrtchyan for discussions on high-derivative equations for higher spins and A. Sagnotti for useful comments. The present research was supported in part by Scuola Normale Superiore, by INFN (I.S. TV12) and by the MIUR-PRIN contract 2009-KHZKRX.


\end{document}